\documentclass[letter, reprint,showpacs,amsmath,amssymb,superscriptaddress]{revtex4-1}
\usepackage{dcolumn}
\usepackage[dvips]{graphicx}
\usepackage{amsmath}
\usepackage{multirow}
\usepackage{setspace}
\newcommand\T{\rule{0pt}{3.1ex}}
\begin{document}

\title{Phase space factors for $\beta^+\beta^+$ decay and competing modes of double-$\beta$ decay} 

\author{J. Kotila}
\email{jenni.kotila@yale.edu}
\affiliation{Center for Theoretical Physics, Sloane Physics Laboratory, Yale University, New Haven, Connecticut, 06520-8120, USA}
\author{F. Iachello}
\email{francesco.iachello@yale.edu}
\affiliation{Center for Theoretical Physics, Sloane Physics Laboratory, Yale University, New Haven, Connecticut, 06520-8120, USA}

\begin{abstract}
A complete and improved calculation of phase space factors (PSF) for $2\nu\beta^+\beta^+$ and $0\nu\beta^+\beta^+$ decay, as well as for the competing modes $2\nu EC\beta^+$, $0\nu EC\beta^+$, and $2\nu ECEC$, is presented. The calculation makes use of exact Dirac wave functions with finite nuclear size and electron screening and includes life-times, single and summed positron spectra, and angular positron correlations.
\end{abstract}

\maketitle
\section{Introduction}

Double-$\beta$ decay is a process in which a nucleus $(A,Z)$ decays to a nucleus $(A,Z\pm2)$ by emitting two electrons or positrons and, usually, other light particles
\begin{equation}
(A,Z)\rightarrow(A,Z\pm 2) + 2e^{\mp} + \text{anything}.
\end{equation}
Double-$\beta$ decay can be classified in various modes according to the various types of particles emitted in the decay.
For processes allowed by the standard model, i.e. the two neutrino modes: $2\nu\beta\beta$, $2\nu\beta EC$, $2\nu ECEC$, the half-life can be, to a good approximation, factorized in the form
\begin{equation}
\label{2nut}
\left[\tau^{2\nu}_{1/2}\right]^{-1}=G_{2\nu}|M_{2\nu}|^2,
\end{equation}
where $G_{2\nu}$ is a phase space factor and $M_{2\nu}$ the nuclear matrix element. For processes not allowed by the standard model, i.e. the neutrinoless modes: $0\nu\beta\beta$, $0\nu\beta EC$, $0\nu ECEC$, the half-life can be factorized as 
\begin{equation}
\label{0nut}
\left[\tau^{0\nu}_{1/2}\right]^{-1}=G_{0\nu}|M_{0\nu}|^2 \left| f(m_i,U_{ei})\right|^2,
\end{equation}
where $G_{0\nu}$ is a phase space factor, $M_{0\nu}$ the nuclear matrix element and $f(m_i, U_{ei})$ contains physics beyond the standard model through the masses $m_i$ and mixing matrix elements $U_{ei}$ of neutrino species. For both processes, two crucial ingredients are the phase space factors (PSF) and the nuclear matrix elements (NME). Recently, we have initiated a program for the evaluation of both quantities and presented results for $\beta^-\beta^-$ decay \cite{barea09,kotila12,barea12,barea12b}. 
This is the most promising mode for the possible detection of neutrinoless double-$\beta$ decay and thus of a measurement of the absolute neutrino mass scale. However, in very recent years, interest in the double positron decay, $\beta^+\beta^+$, positron emitting electron capture, $EC\beta^+$, and double electron capture, $ECEC$, has been renewed. This is due to the fact that positron emitting processes have interesting signatures that could be detected experimentally \cite{zuber}. With this article we initiate a systematic study of  $\beta^+\beta^+$,  $EC\beta^+$, and  $ECEC$ processes. In particular we present here a calculation of phase space factors (PSF). A calculation of nuclear matrix elements (NME), which are common to all three modes, will be presented in a forthcoming publication \cite{barprep12}.

Estimates of the transitions rates for $\beta^+\beta^+$,  $EC\beta^+$, and  $ECEC$ processes were given by Primakoff and Rosen already in the 50's and 60's \cite{primakoff1,primakoff2}. Haxton and Stephenson \cite{haxton} calculated half-lives for $\beta^+\beta^+$ including relativistic corrections approximately and some non-relativistic calculations were done in the 1980's \cite{jotain, jotain2}. In the 90's, this subject was revisited by Doi and Kotani \cite{doitwoneutrino, doineutrinoless, doi} who also presented a detailed theoretical formulation and tabulated results for selected cases. At the same time, Boehm and Vogel \cite{boehm} gave more comprehensive results, but without a detailed theoretical description. In these papers, results for the PSFs were obtained by approximating the positron wave functions at the nucleus and without inclusion of electron screening. In this article, we take advantage of some recent developments in the numerical evaluation of Dirac wave functions and in the solution of the Thomas-Fermi equation to calculate more accurate phase space factors for double-$\beta^+$ decay, $EC\beta^+$ decay, and double-$EC$  in all nuclei of interest. While in the case of $\beta^-\beta^-$ our results (and corrections) were of particular interest in heavy nuclei, $\alpha Z$ large, where relativistic and screening corrections play a major role, in the case of $\beta^+\beta^+$ our results are of interest in all nuclei, since in this case there is a balance between Coulomb repulsion in the final state which favors light nuclei, $\alpha Z$ small, relativistic corrections, which are large for heavy nuclei, $\alpha Z$ large, and screening corrections, which are large in light nuclei due to the opposite sign of $\beta^+\beta^+$ relative to $\beta^-\beta^-$.
 Studies similar to ours were done for single-$\beta^+$ decay and EC in the 1970's \cite{single, bambynek}.
 
In this article we specifically consider  the following five processes:\\
i) Two neutrino double-positron decay, $2\nu\beta^+\beta^+$:
\begin{equation}
(A,Z)\rightarrow(A,Z-2) + 2e^{+} +2\nu 
\end{equation}
ii) Positron emitting two neutrino electron capture, $2\nu EC\beta^+$:
\begin{equation}
(A,Z)+ e^-\rightarrow(A,Z-2) + e^{+} +2\nu 
\end{equation}
iii) Two neutrino double electron capture, $2\nu ECEC$:
\begin{equation}
(A,Z)+ 2e^-\rightarrow(A,Z-2)  +2\nu 
\end{equation}
iv) Neutrinoless double-positron decay, $0\nu\beta^+\beta^+$:
\begin{equation}
(A,Z)\rightarrow(A,Z-2) + 2e^{+} 
\end{equation}
v) Positron emitting neutrinoless electron capture, $0\nu EC\beta^+$:
\begin{equation}
(A,Z)+ e^-\rightarrow(A,Z-2) + e^{+} 
\end{equation}
The neutrinoless double electron capture process $0\nu ECEC$ cannot occur to the order of approximation we are considering, since it must be accompanied by the emission of one or two particles in order to conserve energy, momentum and angular momentum. It will not be considered here.
				
\section{Wave functions}
The key ingredients for the evaluation of phase space factors in single- and double-$\beta$ decay are the scattering wave functions and for EC the bound state wave functions. The general theory of relativistic electrons and positrons can be found e.g., in  the book of Rose \cite{rose}. The electron scattering wave functions of interest in $\beta^-\beta^-$ were given in Eq. (8) of~\cite{kotila12}. In this article, we need the positron scattering wave functions, and the electron bound state wave functions. 

\subsection{Positron scattering wave functions}
We use, for $\beta^+$ decay,  negative energy Dirac central field scattering state wave functions,
\begin{equation}
\psi_{\epsilon\kappa\mu}(\mathbf{r})=\left(
\begin{array}{c}if_{\kappa}(\epsilon,r)\chi_{-\kappa}^{-\mu}\\
-g_{\kappa}(\epsilon,r)\chi_{\kappa}^{-\mu},
\end{array}
\right),
\end{equation}
where $\chi_{\kappa}^{-\mu}$ are spherical spinors and $g_{\kappa}(\epsilon,r)$ and $f_{\kappa}(\epsilon,r)$ are radial functions, with energy $\epsilon$, depending on the relativistic quantum number $\kappa$ defined by $\kappa=(l-j)(2j+1)$. Given an atomic potential $V(r)$ the functions $g_{\kappa}(\epsilon,r)$ and $f_{\kappa}(\epsilon,r)$ satisfy the radial Dirac 
equations:
\begin{equation}
\begin{split}
\frac{dg_{\kappa}(\epsilon,r)}{dr}&=-\frac{\kappa}{r}g_{\kappa}(\epsilon,r)+\frac{\epsilon-V+m_ec^2}{c\hbar}f_{\kappa}(\epsilon,r), \\
\frac{df_{\kappa}(\epsilon,r)}{dr}&=-\frac{\epsilon-V-m_ec^2}{c\hbar}g_{\kappa}(\epsilon,r)+\frac{\kappa}{r}f_{\kappa}(\epsilon,r). 
\end{split}
\end{equation}
The potential $V$ appropriate for this case is obtained from that for electrons by changing the sign of $V$ ($Z$ into $-Z$). These scattering positron wave functions are normalized as the corresponding scattering electron wave functions, Eq. (12) of \cite{kotila12}, except for the change in sign in the Sommerfeld parameter $\eta=Ze^2/\hbar v$. 

\subsection{Electron bound wave functions}
For electron capture (EC) we use positive energy Dirac central field bound state wave functions,
\begin{equation}
\psi_{n`\kappa\mu}(\mathbf{r})=\left(
\begin{array}{c}
g^b_{n`,\kappa}(r)\chi_{\kappa}^{\mu}\\
if^b_{n`,\kappa}(r)\chi_{-\kappa}^{\mu},
\end{array}
\right),
\end{equation}
where $n'$ denotes the radial quantum number and the quantum number $\kappa$ is related to the total angular momentum, $j_{\kappa}=|\kappa|-1/2$. For $K$-shell electrons $n'=0$, $\kappa=-1$, $1S_{1/2}$, while for $L_I$-shell electrons  $n'=1$, $\kappa=-1$, $2S_{1/2}$. We do not consider here $L_{II}$ and $L_{III}$-shells because these are suppressed by the non-zero orbital angular momentum, $2P_{1/2}$, $2P_{3/2}$. The bound state wave functions are normalized in the usual way
\begin{equation}
\begin{split}
&\int \psi_{n`\kappa\mu}(\mathbf{r})^{\dagger}\psi_{n`\kappa\mu}(\mathbf{r})\mathrm{d}\mathbf{r}=\\
&\int_0^\infty \left[ {g^b_{n`,\kappa}}^2(r)+{f^b_{n`,\kappa}}^2(r) \right] \mathrm{d}r=1.
\end{split}
\end{equation}

\subsection{Potential}
The radial positron scattering and electron bound wave functions are evaluated by means of the
subroutine package RADIAL \cite{sal95}, which implements a robust solution method that avoids the accumulation of truncation errors. This is done by solving the radial equations by using a piecewise exact power series expansion of the radial functions, which then are summed up to the prescribed accuracy so that truncation errors can be completely avoided. The input in the package is the potential $V$.  This potential is primarily the Coulomb potential of the daughter nucleus with charge $Z_d$, $V(r)=Z_d(\alpha\hbar c)/r$ in case of $\beta^+$ decay and the Coulomb potential of the mother nucleus with charge $Z_m$, $V(r)=-Z_m(\alpha\hbar c)/r$ in case of electron capture. As in the case of single-$\beta$ decay and electron capture we include nuclear size corrections and screening.

The nuclear size corrections are taken into account by an uniform charge distribution  in a sphere of radius $R=r_0A^{1/3}$ with $r_0=1.2$ fm, i.e. 
\begin{equation}
V(r)=\left[\begin{array}{lll}
\pm\frac{Z_{i}(\alpha\hbar c)}{r}&, &r\geq R\\
\pm Z_{i}(\alpha\hbar c)\left(\frac{3-(r/R)^2}{2R}\right)&, &r< R
\end{array}\right], 
\end{equation}
$i=d,m$. The introduction of finite nuclear size has also the advantage that the singularity at the origin in the solution of the Dirac equation is removed.

\begin{figure}[ht!]
\includegraphics[width=1.\linewidth ]{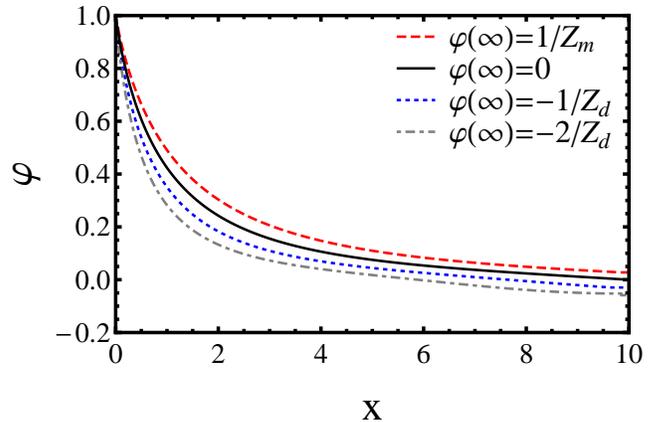} 
\caption{\label{tffig}(Color online) The Thomas-Fermi functions with the boundary conditions of Eqs. (\ref{bound1}) -(\ref{bound3}) for $^{78}$Kr decay. The dotdashed (gray) curve corresponds to the solution for $\beta^+\beta^+$ decay, the dashed (red) curve corresponds to the solution for EC, the dashed (blue) curve corresponds to the solution for $\beta^+$ decay, and the solid (black) curve corresponds to the ECEC.}
\end{figure}

The contribution of screening to the phase space factors was extensively investigated in single-$\beta$ decay \cite{wil70, buhring}. The screening potential is of order $V_S\propto Z_i^{4/3}\alpha^2$ and thus gives a contribution of order $\alpha=1/137$ relative to the pure Coulomb potential $V_C\propto Z_i \alpha$. 
We take the screening contribution  into account by using the Thomas-Fermi approximation.  The Thomas-Fermi function $\varphi(x)$, solution of the Thomas-Fermi equation
\begin{equation}
\label{tf}
\frac{d^2\varphi}{dx^2}=\frac{\varphi^{3/2}}{\sqrt{x}}
\end{equation}
with $x=r/b$  and
\begin{equation}
b=\frac{1}{2}\left(\frac{3\pi}{4}\right)^{2/3}\frac{\hbar ^2}{m_ee^2}Z_i^{-1/3}\simeq 0.8853a_0Z_i^{-1/3},
\end{equation}
where $i=d,m$ and $a_0$ is the Bohr radius, is obtained by solving Eq.~(\ref{tf}) for a point charge $Z_i$ with boundary conditions
\begin{equation}
\begin{split}
\label{bound1}
\varphi(0)&=1,\\
\varphi(\infty)&=-\frac{2}{Z_d}, 
\end{split}
\end{equation} 
for $\beta^+\beta^+$ decay,
\begin{equation}
\begin{split}
\label{bound2}
\varphi(0)&=1,\\
\varphi(\infty)&=\frac{1}{Z_m}, \hspace{1cm}(EC)\\
\varphi(\infty)&=-\frac{1}{Z_d},\hspace{0.9cm} (\beta^+) 
\end{split}
\end{equation} 
for $EC\beta^+$ decay ($EC$, $\beta^+$, respectively), and
\begin{equation}
\begin{split}
\label{bound3}
\varphi(0)&=1,\\
\varphi(\infty)&=0, 
\end{split}
\end{equation} 
for $ECEC$ decay.
This takes into account the fact that the final atom is a negative ion with charge $-2$, $-1$ or a neutral ion depending on the mode ($\beta^+\beta^+$, $EC\beta^+$, $ECEC$, respectively). With the introduction of this function, the potential $V(r)$ including screening becomes
\begin{equation}
V(r)\equiv \varphi(r) \times
\left[\begin{array}{lll}
\pm \frac{Z_i(\alpha\hbar c)}{r} &, &r\geq R\\
\pm Z_i(\alpha\hbar c)\left(\frac{3-(r/R)^2}{2R}\right) &, &r<R
\end{array}\right],
\end{equation}
$i=d,m$. This can be rewritten in terms of an effective charge $Z_{\rm{eff}}=Z_i\varphi(r)$ where $Z_{\rm{eff}}$ now depends on $r$. In order to solve Eq.~(\ref{tf}), we use the Majorana method described in \cite{esp02} which is valid both for a neutral atom and negative/positive ion. The Majorana method requires only one quadrature and is amenable to a simple solution, the accuracy of which depends on the number of terms kept in the series expansion of the auxiliary function $u(t)$ of Ref.~\cite{esp02}. The solution is smooth for all three boundary conditions. It is particularly useful here, since we want to evaluate screening corrections in several nuclei. As an example for the resulting $\varphi(x)$ functions with the boundary conditions presented in Eqs.~(\ref{bound1}) -(\ref{bound3}) we show in Fig.~\ref{tffig} results for $^{78}$Kr decay.

\subsection{Solutions}
\begin{figure*}[cbt!]
\includegraphics[width=0.85\linewidth ]{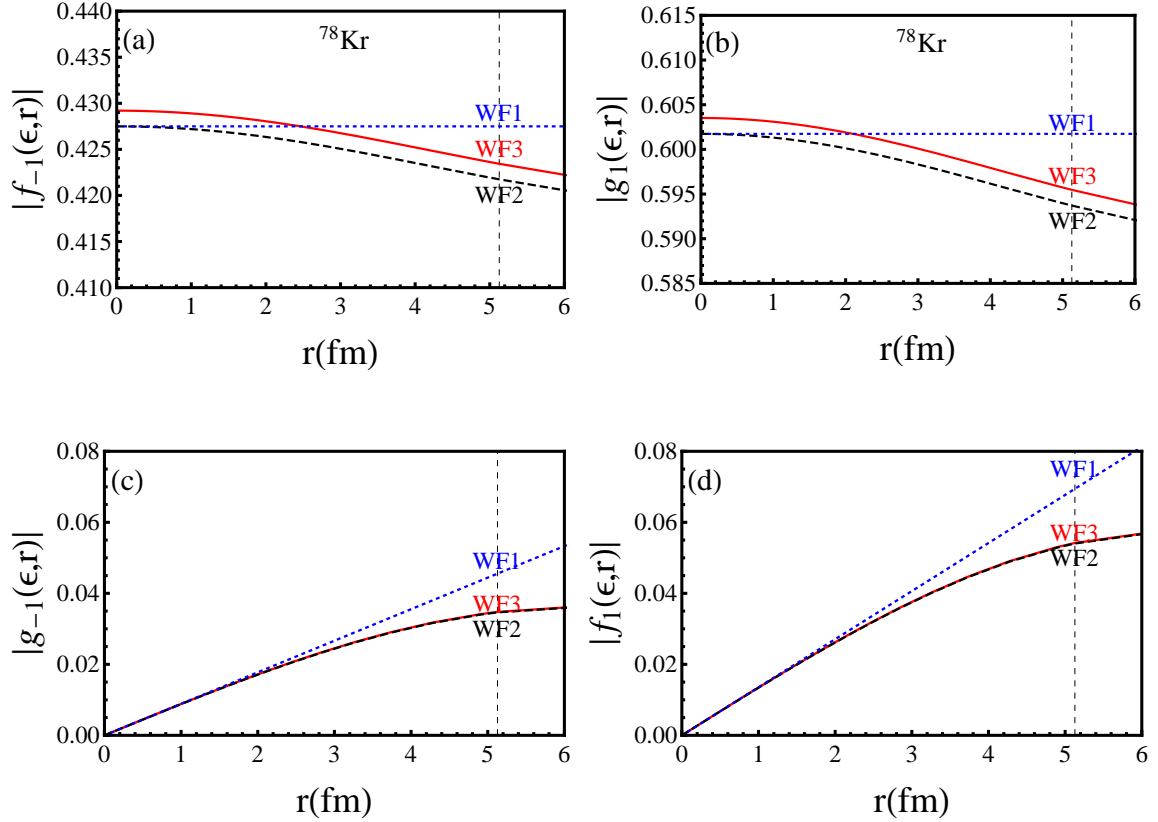} 
\caption{\label{poswave}(Color online) Positron radial wave functions $f_{-1}(\epsilon,r)$, $g_{-1}(\epsilon,r)$ (panels a and c,  respectively) and $g_{1}(\epsilon,r)$, $f_{1}(\epsilon,r)$ (panels b and d, respectively) for $Z_d=34$, $\epsilon=1.0$ MeV and $R=5.13$ fm (vertical line). The notations WF1 (dotted lines), WF2 (dashed lines), and WF3 (solid lines) correspond to leading finite size Coulomb, exact finite size Coulomb and  exact finite size Coulomb with electron screening, respectively.}
\end{figure*}

\begin{figure*}[cbt!]
\includegraphics[width=0.85\linewidth ]{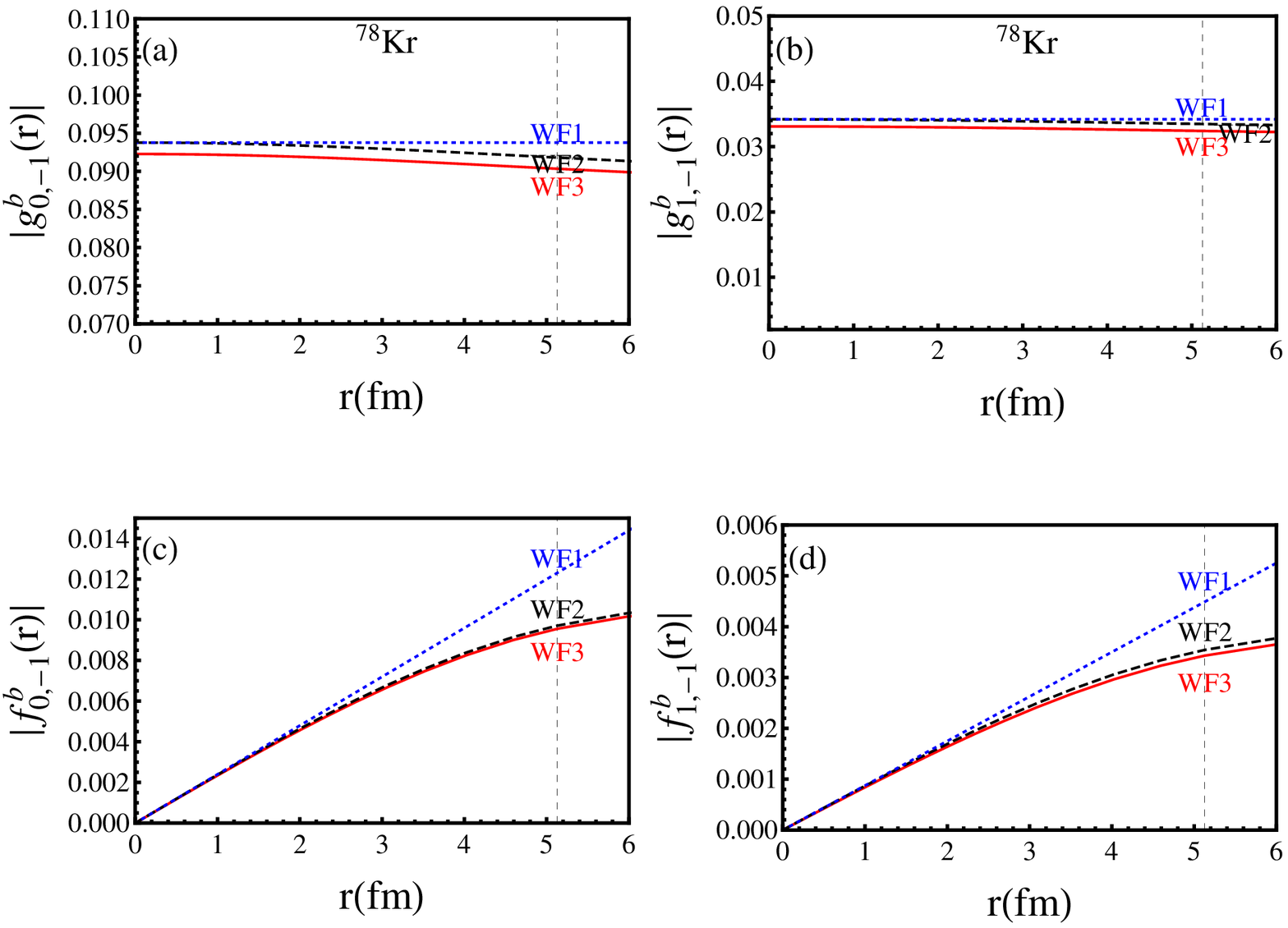} 
\caption{\label{boundwave}(Color online) Electron bound state wave functions $g_{0,-1}^b(r)$, $f_{0,-1}^b(r)$ (panels a and c,  respectively) and $g_{1,-1}^b(r)$, $f_{1,-1}^b(r)$ (panels b and d, respectively) for $Z_m=36$ and $R=5.13$ fm (vertical line) scaled dimensionless with a factor of $[4\pi(m_ec^2)^3]^{-1/2}(\hbar c/a_0)^{3/2}a_0$. The notations WF1 (dotted lines), WF2 (dashed lines), and WF3 (solid lines) correspond to leading finite size Coulomb, exact finite size Coulomb and  exact finite size Coulomb with electron screening, respectively.}
\end{figure*}

In order to illustrate the effect of finite size and screening we show in Fig.~\ref{poswave} the positron scattering wave function for $\epsilon=1.0$MeV, and in Fig.~\ref{boundwave} the electron bound wave function for the $1S_{1/2}$ and $2S_{1/2}$ states. Comparing Fig.~\ref{boundwave} with  Fig. 2 of Ref.~\cite{kotila12} Fig. 2, one can see that the effect of screening is larger than in $\beta^-\beta^-$ and of opposite sign, since the electron cloud decreases the magnitude of the repulsive potential seen by the outgoing positrons.

\section{Phase space factors in double-$\beta$ decay}
In order to calculate PSFs for $\beta^+\beta^+$, $EC\beta^+$, and $ECEC$, we use the formulation of Doi and Kotani \cite{doitwoneutrino, doineutrinoless}.
\subsection{Decays where two neutrinos are emitted}
\begin{figure}[ht!]
\includegraphics[width=1.00\linewidth ]{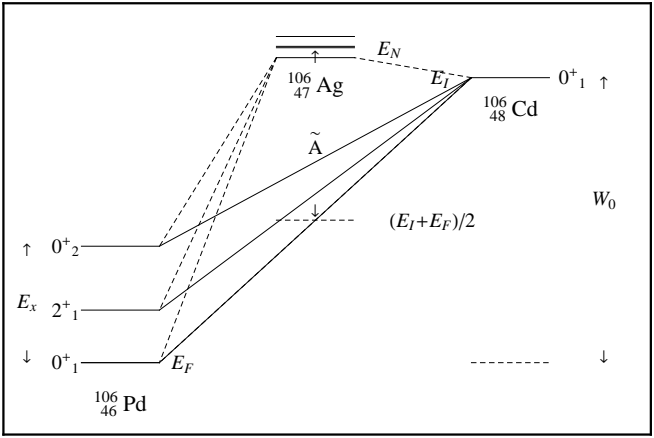} 
\caption{\label{scheme}Notation used in this article. The example is for $^{106}$Cd decay.}
\end{figure}

The $2\nu\beta\beta$ decay is a second order process in the effective weak interaction. It can be calculated in a way analogous to single-$\beta$ decay. 
Neglecting the neutrino mass, considering only S-wave states and noting that with four leptons in the final state we can have angular momentum $0$, $1$ and, $2$, we see that both $0^+\rightarrow 0^+$ and $0^+\rightarrow 2^+$ decays can occur. We denote by $Q_i$, where $i=\beta^+\beta^+, EC\beta^+, ECEC$,  the $Q$-values of the decay. These can be obtained from the mass difference between neutral mother and daughter atoms, $M(A,Z)-M(A,Z-2)$ as
\begin{equation}
\begin{split}
Q_{\beta^+\beta^+}&=M(A,Z)-M(A,Z-2)-4m_ec^2,\\
Q_{EC\beta^+}&=M(A,Z)-M(A,Z-2)-2m_ec^2,\\
Q_{ECEC}&=M(A,Z)-M(A,Z-2).\\
\end{split}
\end{equation}
For the total available kinetic energy one also needs to take into account the binding energy of the captured electron and thus the total available kinetic energies for $\beta^+\beta^+$, $EC\beta^+$ and $ECEC$ modes are
\begin{equation}
\begin{split}
T_{\beta^+\beta^+}&=M(A,Z)-M(A,Z-2)-4m_ec^2,\\
T_{EC\beta^+}&=M(A,Z)-M(A,Z-2)-2m_ec^2-\epsilon_b\\
T_{ECEC}&=M(A,Z)-M(A,Z-2)-\epsilon_{b_1}-\epsilon_{b_2}.\\
\end{split}
\end{equation}
The values of $M(A,Z)-M(A,Z-2)$ are shown in Table~\ref{Qval}. Another quantity of interest in the evaluation of the PSFs is the excitation energy $E_N$ of the intermediate nucleus with respect to the average of the initial and final ground states,
\begin{equation}
\begin{split}
\tilde{A}=\frac{1}{2} &W_0+ E_N-E_I=\frac{1}{2}[M(A,Z) \\
&  - M(A,Z-2)-2m_e c^2]+ E_N-E_I,
\end{split}
\end{equation}
illustrated in Fig.~\ref{scheme}. As discussed in Ref.~\cite{kotila12}, the results for PSF depend weakly on the values of the energies $E_N$ in the intermediate odd-odd nucleus, as remarked years ago by Tomoda \cite{tom91} and as shown explicitly in our Ref.~\cite{kotila12}, Fig.~4. We therefore perform all calculations in this paper by replacing $E_N$ with an average value $\langle E_N \rangle$ and $\tilde{A}=1.12 A^{1/2}$MeV as suggested by Haxton and Stephenson \cite{haxton}. The error introduced by this approximation is discussed in the following Sect.~\ref{secterr}. We emphasize, however, that our calculation has been set up in such a way as to allow a state by state evaluation, if needed. 

\begin{ruledtabular}
\begin{center}
\begin{table}[h!]
\begin{tabular}{lc}
Nucleus   &$M(A,Z)-M(A,Z-2)$(MeV)\footnotemark[1] \cr \hline
 \multicolumn{2}{c}{\T \textbf{$\beta^+\beta^+$, $EC\beta^+$ and $ECEC$ allowed}}\\
$^{78}$Kr	&2.8463(7) \\
$^{96}$Ru	&2.71451(13)\footnotemark[2] \\
$^{106}$Cd	&2.77539(10)\footnotemark[3] \\
$^{124}$Xe	&2.8654(22) \\
$^{130}$Ba	&2.619(3) \\
$^{136}$Ce	&2.37853(27)\footnotemark[4] \\
\multicolumn{2}{c}{\T \textbf{$EC\beta^+$ and $ECEC$ allowed}}\\		
$^{50}$Cr	&1.1688(9) \\
$^{58}$Ni	&1.9263(3) \\
$^{64}$Zn	&1.0948(7) \\
$^{74}$Se	&1.209169(49)\footnotemark[5] \\
$^{84}$Sr	&1.7900(13) \\
$^{92}$Mo	&1.651(4) \\
$^{102}$Pd	&1.1727(36)\footnotemark[3] \\
$^{112}$Sn	&1.91982(16)\footnotemark[6] \\
$^{120}$Te	&1.71481(125)\footnotemark[7] \\
$^{144}$Sm	&1.78259(87)\footnotemark[3] \\
$^{156}$Dy	&2.012(6) \\
$^{162}$Er	&1.8440(30)\footnotemark[2] \\
$^{168}$Yb	&1.40927(25)\footnotemark[2] \\
$^{174}$Hf	&1.0988(23) \\
$^{184}$Os	&1.453(58)\footnotemark[8] \\
$^{190}$Pt	&1.384(6) \\
\multicolumn{2}{c}{\T \textbf{$ECEC$ allowed}}\\		
$^{36}$Ar	&0.43259(19) \\
$^{40}$Ca	&0.193510(20)\\
$^{54}$Fe	&0.6798(4) \\
$^{108}$Cd	&0.27204(55)\footnotemark[9] \\
$^{126}$Xe	&0.920(4) \\
$^{132}$Ba	&0.8440(10) \\
$^{138}$Ce	&0.698(10) \\
$^{152}$Gd	&0.05570(18)\footnotemark[10] \\
$^{158}$Dy	&0.284(3) \\
$^{164}$Er	&0.02507(12)\footnotemark[11] \\
$^{180}$W	&0.14320(27)\footnotemark[12] \\
$^{196}$Hg	&0.820(3) \\
\end{tabular}
\footnotetext[1]{Ref.~\cite{nudat}}
\footnotetext[2]{Ref.~\cite{prc83}}
\footnotetext[3]{Ref.~\cite{prc84}}
\footnotetext[4]{Ref.~\cite{plb697}}
\footnotetext[5]{Ref.~\cite{plb684}}
\footnotetext[6]{Ref.~\cite{prl103}}
\footnotetext[7]{Ref.~\cite{prc80}}
\footnotetext[8]{Ref.~\cite{prc86}}
\footnotetext[9]{Ref.~\cite{arxiv1201}}
\footnotetext[10]{Ref.~\cite{prl106}}
\footnotetext[11]{Ref.~\cite{prl107}}
\footnotetext[12]{Ref.~\cite{arxiv111}}
\caption{\label{Qval}Mass difference $M(A,Z)-M(A,Z-2)$ 
 used in the calculation.}
\end{table}
\end{center}
\end{ruledtabular}

\subsubsection{$2\nu\beta^+\beta^+$ decay}
The formulas for $2\nu\beta^+\beta^+$ decay are exactly the same as for $2\nu\beta^-\beta^-$ decay described in \cite{kotila12} where now $\epsilon_1$ is the energy of the first positron, $\epsilon_1=\epsilon_{p_1}$, and $\epsilon_2$ is the energy of the second positron, $\epsilon_2=\epsilon_{p_2}$. We use here the same approximations as in \cite{kotila12}, that is to evaluate the positron wave functions at the nuclear radius
\begin{equation}
\label{I}
\begin{split}
g_{-1}(\epsilon)&=g_{-1}(\epsilon,R)\\
f_{1}(\epsilon)&=f_{1}(\epsilon,R)
\end{split}
,
\end{equation}
and to replace the excitation energy $E_N$ in the intermediate odd-odd nucleus by a suitably chosen energy $\langle E_N \rangle$, giving 
\begin{equation}
\tilde{A}=\frac{1}{2}W_0+\langle E_N \rangle -E_I.
\end{equation}
The phase space factors are then given in terms of quantities \cite{kotila12, tom91}
\begin{equation}
\label{closure}
\begin{split}
\left< K_N\right>=&\frac{1}{\epsilon_1+\omega_1+\left< E_N\right> -E_I}\\
&+\frac{1}{\epsilon_2+\omega_2+\left< E_N\right> -E_I},\\
\left< L_N\right>=&\frac{1}{\epsilon_1+\omega_2+\left< E_N\right> -E_i}\\
&+\frac{1}{\epsilon_2+\omega_1+\left< E_N\right> -E_I}.
\end{split}
\end{equation}
These approximations allow a separation of the PSF from the nuclear matrix elements and the condition under which they are good have been discussed in \cite{kotila12}. Apart from a narrow region around threshold, where the error is $\sim 1\%$, the approximations are good throughout. For $\beta^+\beta^+$ decay we have two integrated phase space factors $G_{2\nu}^{(0)}$ and $G_{2\nu}^{(1)}$ whose explicit expression are given in Eqs.~(21)-(28) and (34)-(36) of \cite{kotila12}. Since the calculated single-$\beta$ decay matrix elements of the GT operator in a particular nuclear model appear to be systematically larger than those derived from measured $ft$ values of the allowed GT transitions, and this effect is usually taken into account by quenching the axial vector coupling constant $g_{A,eff}=qg_A$, it is convenient to separate it from the phase space factors $G_{2\nu}$. Also, it is convenient to scale the matrix elements with the electron mass, $m_ec^2$. The phase space factors are then in units of yr$^{-1}$. From these we obtain\\
(i) The half-life
\begin{equation}
\label{t1/2_2nbb}
\left[ \tau^{2\nu}_{1/2}\right]^{-1}=G^{\beta^+\beta^+}_{2\nu}g_A^4 \left| m_ec^2 M^{(2\nu)} \right| ^2.
\end{equation}
(ii) The differential decay rate
\begin{equation}
\frac{dW_{2\nu}}{d\epsilon_{p_1}}={\cal N}_{2\nu}\ln 2\frac{dG^{\beta^+\beta^+}_{2\nu}}{d\epsilon_{p_1}},
\end{equation}
where ${\cal N}_{2\nu}=g_A^4 \left| m_ec^2 M^{(2\nu)} \right|^2$.\\
(iii) The summed energy spectrum of the two positrons
\begin{equation}
\frac{dW_{2\nu}}{d(\epsilon_{p_1}+\epsilon_{p_2})}={\cal N}_{2\nu} \ln 2\frac{dG^{\beta^+\beta^+}_{2\nu}}{d(\epsilon_{p_1}+\epsilon_{p_2})}.
\end{equation}
These three quantities depend only on $G_{2\nu}^{(0)}\equiv G_{2\nu}^{\beta^+\beta^+}$.\\
(iv) The angular correlation between the two positrons
\begin{equation}
\alpha(\epsilon_1)=\frac{dG^{(1)}_{2\nu}/d\epsilon_{p_1}}{dG^{(0)}_{2\nu}/d\epsilon_{p_1}},
\end{equation}
which depend on both $G_{2\nu}^{(0)}$ and $G_{2\nu}^{(1)}$. Here and in the following subsections 2 and 3,
\begin{equation}
M^{(2\nu)}= -\left[\frac{M_{GT}^{(2\nu)}}{\tilde{A}_{GT}}-\left(\frac{g_{V}}{g_{A}}\right)^{2}\frac{M_{F}^{(2\nu)}}{ \tilde{A}_{F}}\right],
\end{equation}
where $M_{GT}^{(2\nu)} =  \left\langle 0_{F}^{+}\left\vert \sum_{nn^{\prime}}\tau_{n}^{\dag}\tau_{n^{\prime}}^{\dag}\vec{\sigma}_{n}\cdot\vec{\sigma}_{n^{\prime}}\right\vert 0_{I}^{+}\right\rangle$ and $M_{F}^{(2\nu)}  =  \left\langle 0_{F}^{+}\left\vert \sum_{nn^{\prime}}\tau_{n}^{\dag}\tau_{n^{\prime}}^{\dag}\right\vert 0_{I}^{+}\right\rangle$.
The closure energies $\tilde{A}_{GT}$ and $\tilde{A}_{F}$ could in principle be different, but in this article we take $\tilde{A}_{GT}=\tilde{A}_{F} \equiv \tilde{A}$.

\begin{figure}[ht!]
\includegraphics[width=1.00\linewidth ]{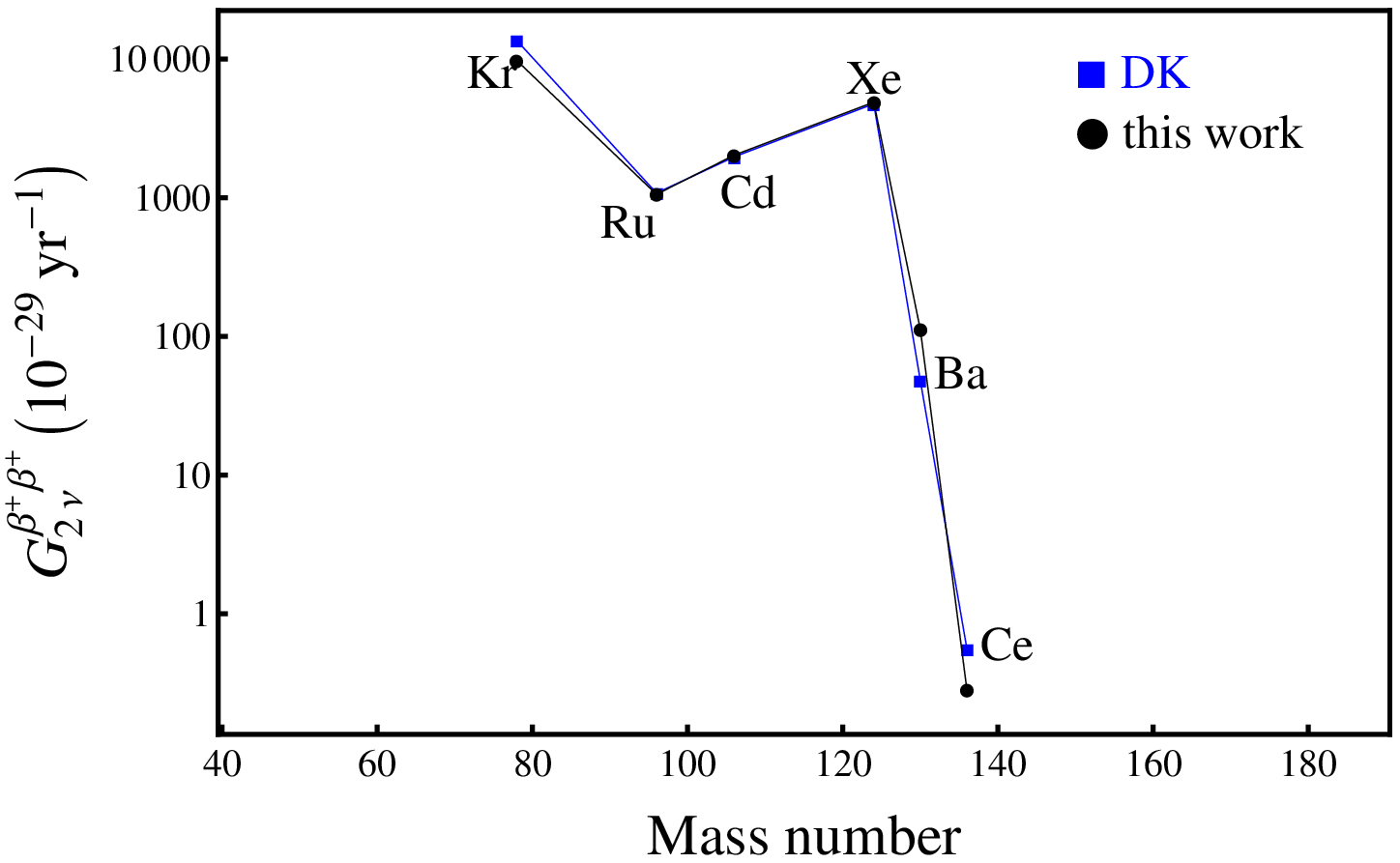} 
\caption{\label{2nuGcompbb}(Color online) Phase space factors $G_{2\nu}^{\beta^+\beta^+}$ in units $(10^{-29}$ yr$^{-1})$. The label "DK" refers to the results obtained by Doi and Kotani \cite{doitwoneutrino} using approximate electron wave functions. The figure is in semilogarithmic scale.}
\end{figure}

\begin{figure*}[cbt!]
\includegraphics[width=1.00\linewidth ]{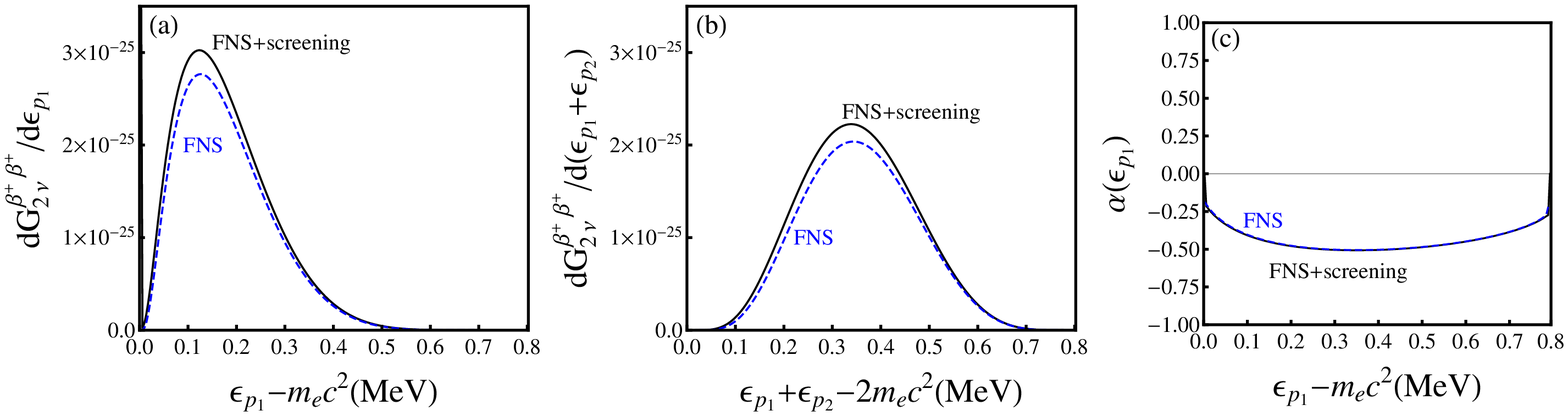} 
\caption{\label{kr}(Color online) Single positron spectra (panel a), summed energy spectra (panel b) and angular correlations between the two outgoing positrons (panel c) for the $^{78}$Kr $\rightarrow ^{78}$Se $2\nu\beta^+\beta^+$-decay. The scale in the left and middle panels should be multiplied by ${\cal N}_{2\nu}$ when comparing with experiment. In panels a and b the upper, solid curve is obtained when taking into account finite nuclear size and electron screening, while the lower, dashed curve presents spectra obtained when taking into account only the finite nuclear size. In panel c these two calculations coincide.}
\end{figure*}

\begin{figure*}[cbt!]
\includegraphics[width=1.0\linewidth ]{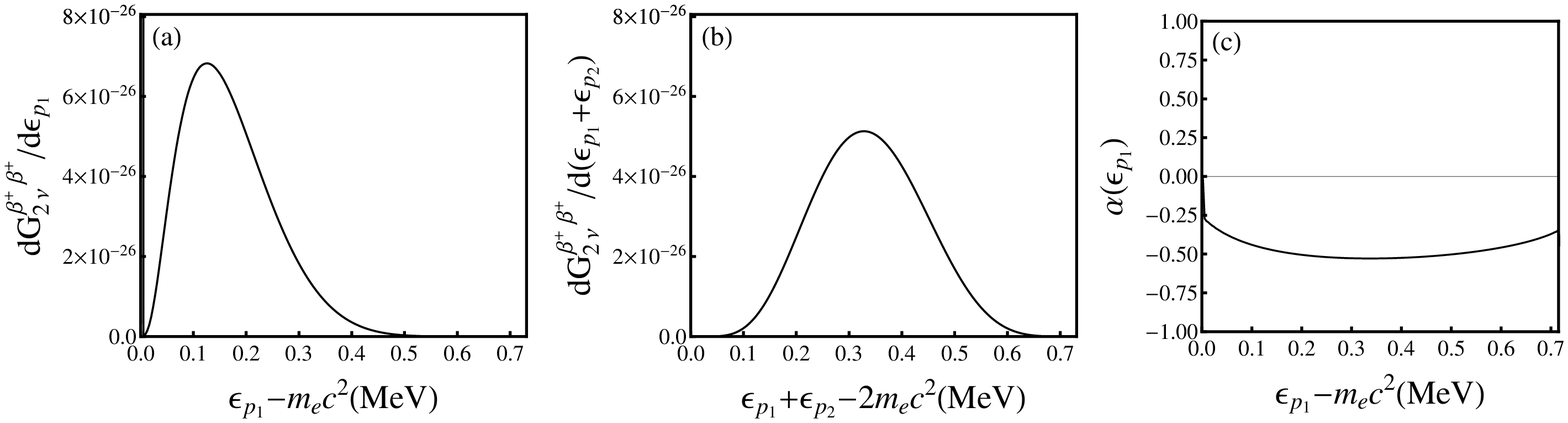} 
\caption{\label{cd}Single positron spectra (panel a), summed energy spectra (panel b) and angular correlations between the two outgoing positrons (panel c) for the $^{106}$Cd $\rightarrow ^{106}$Pd $2\nu\beta^+\beta^+$-decay. The scale in the left and middle panels should be multiplied by ${\cal N}_{2\nu}$ when comparing with experiment.}
\end{figure*}

The phase space factors for $2\nu\beta^+\beta^+$ decay are listed in Table~\ref{table2} column 2, where they are also compared with values found from literature \cite{doitwoneutrino, boehm} (columns 3 and 4), and in Fig.~\ref{2nuGcompbb}. The values in the literature have been converted to our notation by removing factors of $g_A^4$ and $(m_ec^2)^2$. The value for $^{136}$Ce should be taken with caution because of the very low $Q$-value. We also have available upon request for all $2\nu\beta^+\beta^+$ nuclei in Table~\ref{table2} the single positron spectra, the summed energy spectra and angular correlations between the two outgoing positrons. As examples, we show the cases of $^{78}$Kr~$\rightarrow~^{78}$Se decay, Fig.~\ref{kr}, 
 and of  $^{106}$Cd~$\rightarrow~^{106}$Pd, Fig.~\ref{cd}. 
  The use of a screened potential makes a considerable difference compared to the results obtained when taking into account only the finite nuclear size, as shown in Fig.~\ref{kr}. 
 Note the difference between the single positron spectra in Figs.~\ref{kr} and \ref{cd} for $\beta^+\beta^+$ and the single electron spectra in Figs. 6 and 7 of \cite{kotila12} for $\beta^-\beta^-$decay. Note also the difference in the scale of Fig.~\ref{2nuGcompbb}, $10^{-29}$yr$^{-1}$, for $2\nu\beta^+\beta^+$, as compared with the scale of Fig. 5 of \cite{kotila12}, $10^{-21}$yr$^{-1}$, for $2\nu\beta^-\beta^-$.
\begin{ruledtabular}
\begin{center}
\begin{table*}[cbt!]
\begin{tabular}{lccccccccc}

   		&\multicolumn{3}{c}{$G_{2\nu}^{\beta^+\beta^+}(10^{-29}yr^{-1})$}		&\multicolumn{3}{c}{$G_{2\nu}^{EC\beta^+}(10^{-24}yr^{-1})$}		&\multicolumn{3}{c}{$G_{2\nu}^{ECEC}(10^{-24}yr^{-1})$}	\\\cline{2-4} \cline{5-7} \cline{8-10}
 \T
Nucleus   	&This work &DK & BV &This work &DK &BV &This work &DK &BV\\
\hline
\T&\multicolumn{9}{c}{ \textbf{$\beta^+\beta^+$, $EC\beta^+$ and $ECEC$ allowed}}\\
$^{78}$Kr	&9770	&13600&16000	&385 &464&390		&660&774 &136\\	
$^{96}$Ru	&1040	&1080&1230		&407 &454&350		&2400&2740&433\\	
$^{106}$Cd	&2000	&1970&2420		&702 &779&652		&5410&6220&1120\\	
$^{124}$Xe	&4850	&4770&5410		&1530 &1720&1408	&17200&20200&3500\\	
$^{130}$Ba	&110	&47.9&59.2		&580 &549&420		&15000&16300&2590\\	
$^{136}$Ce	&0.267	&0.559&0.795	&190 &253&192		&12500&15800&2420\\	
\T &\multicolumn{9}{c}{ \textbf{$EC\beta^+$ and $ECEC$ allowed}}\\								

$^{50}$Cr		&		&	&&1.16$\times 10^{-6}$	&	&1.05$\times 10^{-6}$	&0.422&&0.0887\\	
$^{58}$Ni		&		&	&&1.11	&1.16&1.00	 		&15.3&17.0&3.01\\	
$^{64}$Zn		&		&	&&3.81$\times 10^{-9}$	&	&3.83$\times 10^{-9}$	&1.41&&0.281\\	
$^{74}$Se		&		&	&&1.09$\times 10^{-5}$	&	&8.39$\times 10^{-6}$	&5.656&&1.08\\	
$^{84}$Sr		&		&	&&0.729	&	&0.616		&93.6&&17.9\\	
$^{92}$Mo		&		&	&&0.206	&	&0.164		&208&&24.0\\	
$^{102}$Pd		&		&	&&1.62$\times 10^{-6}$	&	&7.16$\times 10^{-7}$		&46.0&&9.14\\	
$^{112}$Sn		&		&	&&4.95	&	&4.33		&1150&&235\\	
$^{120}$Te		&		&	&&0.730	&	&0.524		&888&&173\\	
$^{144}$Sm		&		&	&&2.49	&	&1.98		&5150&&982\\	
$^{156}$Dy		&		&	&&25.3	&	&20.2		&17600&&3100\\	
$^{162}$Er		&		&	&&6.40	&6.69&5.29		&15000&18100&2770\\	
$^{168}$Yb		&		&	&&0.00979	&	&0.00763	&4710&&890\\	
$^{174}$Hf		&		&	&&1.00$\times 10^{-9}$	&	&2.77$\times 10^{-14}$	&1580&&310\\	
$^{184}$Os		&		&	&&0.0299	&	&0.0156   	&12900&&2240\\	
$^{190}$Pt		&		&	&&0.00588	&	&0.00235	&12900&&2290\\

\T &\multicolumn{9}{c}{ \textbf{$ECEC$ allowed}}\\
							

$^{40}$Ca	&&			&		&		&&&1.25$\times 10^{-5}$&&\\	
$^{54}$Fe	&&			&		&		&&&0.0469&&\\	
$^{108}$Cd	&&			&		&		&&&0.0207&&\\	
$^{126}$Xe	&&			&		&		&&&46.1&&\\	
$^{132}$Ba	&&			&		&		&&&39.1&&\\	
$^{138}$Ce	&&			&		&		&&&18.4&&\\	
$^{158}$Dy	&&			&		&		&&&0.183&&\\	
$^{180}$W	&&			&		&		&&&0.00156&&\\	
$^{196}$Hg	&&			&		&		&&&821&&\\	
\end{tabular}
\caption{\label{table2}Phase space factors $G_{2\nu}$ obtained using screened exact finite size Coulomb wave functions. For comparison, values of Doi and Kotani \cite{doitwoneutrino} and Boehm and Vogel \cite{boehm} are also shown. They have been extracted from \cite{doitwoneutrino} and \cite{boehm} by removing $g_A^4$, and converting to yr$^{-1}$ units.}
\end{table*}
\end{center}
\end{ruledtabular}

\subsubsection{$2\nu EC\beta^+$ decay}

For the calculation of electron capture processes the crucial quantity is the probability that an electron is found at the nucleus. This can be expressed in terms of the dimensionless quantity \cite{doitwoneutrino}
\begin{equation}
\begin{split}
{\cal B}_{n',\kappa}^2=&\frac{1}{4\pi(m_ec^2)^3}\left( \frac{\hbar c}{a_0}\right)^3 \left( \frac{a_0}{R}\right) ^2 \\
&\times \left[ \left(g_{n',\kappa}^b(R)\right)^2 +\left(f_{n',\kappa}^b(R)\right)^2 \right],
\end{split}
\end{equation}
where $a_0$ is the Bohr radius $a_0=0.529 \times 10^{-8}$cm and we use for the nuclear radius $R=1.2A^{1/3}$fm. For capture from the $K$-shell $n'=0$, $\kappa=-1$, $1S_{1/2}$ while for capture from the $L_I$-shell $n'=1$, $\kappa=-1$, $2S_{1/2}$.

Denoting by $\epsilon_p$ the energy of the emitted positron and by $e_b$ the binding energy of the captured electron, the phase space factor can be written as \cite{doitwoneutrino}
\begin{small}
\begin{equation}
\begin{split}
G^{EC\beta^+}_{2\nu}&=\frac{2\tilde{A}^2}{3\ln2}\frac{(G\cos\theta)^4}{16\pi^5 \hbar}(m_ec^2) \sum_{i=0,1}{\cal B}_{i,-1}^2 \int^{ Q_{EC\beta^+}+\epsilon_b +m_ec^2}_{m_ec^2}
\\&\times 
\int^{ Q_{EC\beta^+}+\epsilon_b -\epsilon_p}_{0}
\left[ (g_{-1}(\epsilon_p ,R))^2+(f_{1}(\epsilon_p ,R))^2\right]\\
&\times \left( \left< K_N \right> ^2 +\left< L_N \right> ^2 +\left< K_N \right> \left< L_N \right> \right)
\omega_1^2 \omega_2^2 p_pc \epsilon_p d\omega_1 d\epsilon_p,
\end{split}
\end{equation}
\end{small}
where $\omega_1$ and $\omega_2$ are the neutrino energies.
Now in the definition of $\left< K_N \right>$ and $\left< L_N \right>$ in Eq. (\ref{closure}), $\epsilon_1=\epsilon_e=-(m_ec^2-e_b)$ is the energy of the captured electron and $\epsilon_2=\epsilon_p$ is the energy of emitted positron.
Again separating $g_A^4$ and the electron mass $(m_ec^2)^2$, the PSF are in units of yr$^{-1}$. From those, we obtain:\\
(i) The half-life
\begin{equation}
\label{t1/2_2necb}
\left[ \tau^{2\nu}_{1/2}\right]^{-1}=G^{EC \beta^+}_{2\nu}g_A^4 \left| m_ec^2 M^{(2\nu)} \right| ^2.
\end{equation}
(ii) The differential decay rate
\begin{equation}
\frac{dW_{2\nu}}{d\epsilon_p}={\cal N}_{2\nu}\ln 2 \frac{dG^{EC\beta^+}_{2\nu}}{d\epsilon_p},
\end{equation}
where ${\cal N}_{2\nu}=g_A^4 \left| m_ec^2 M^{(2\nu)} \right|^2$.\\
\begin{figure}[ht!]
\includegraphics[width=1.00\linewidth ]{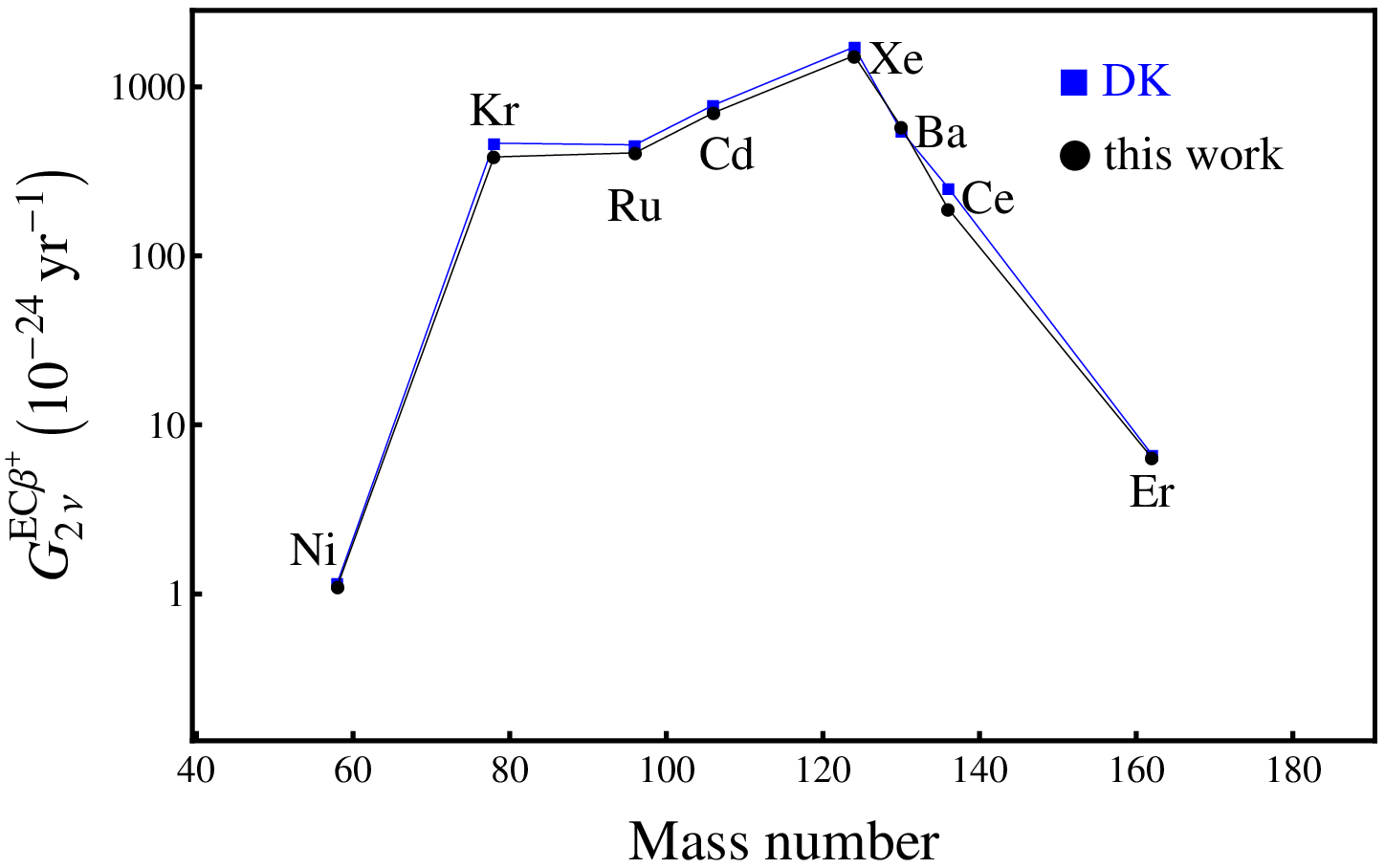} 
\caption{\label{2nuecb}(Color online) Phase space factors $G_{2\nu}^{EC\beta^+}$ in units $(10^{-24}$ yr$^{-1})$. The label "DK" refers to the results obtained by Doi and Kotani \cite{doitwoneutrino} using approximate electron wave functions. The figure is in semilogarithmic scale.}
The obtained PSFs are listed in Table~\ref{table2} column 5 where they are compared with previous calculations (columns 6 and 7), and in Fig.~\ref{2nuecb}. The very small values in the second part of the table should be taken with caution in view of their very small $Q$-value.

An example of single positron spectrum is shown in Fig.~\ref{ni}. This figure is for $^{106}$Cd~$\rightarrow ^{106}$Pd~$2\nu EC\beta^+$-decay.
\end{figure}
\begin{figure}[ht!]
\includegraphics[width=0.81\linewidth ]{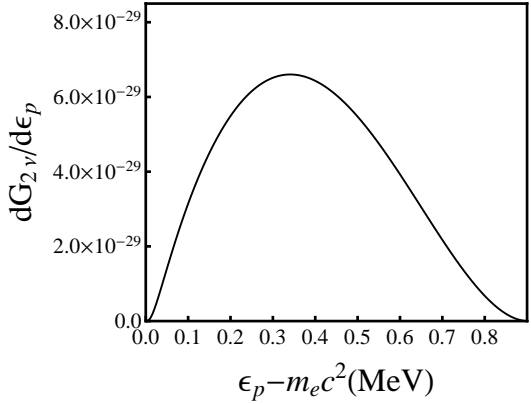} 
\caption{\label{ni}Single positron spectra  for the $^{106}$Cd $\rightarrow ^{106}$Pd $2\nu EC\beta^+$-decay. The scale  should be multiplied by ${\cal N}_{2\nu}$ when comparing with experiment.}
\end{figure}

\subsubsection{$2\nu ECEC$ decay}
In the case of double electron capture with two neutrinos the energies of the electrons are fixed and the two neutrinos carry all the excess energy. The equation for PSF then reads \cite{doitwoneutrino}: 
\begin{widetext}
\begin{equation}
G^{ECEC}_{2\nu}=\frac{2\tilde{A}^2}{3\ln2}\frac{(G\cos\theta)^4}{16\pi^3 \hbar}(m_ec^2)^4 \sum_{i,j=0,1}{\cal B}_{i,-1}^2{\cal B}_{j,-1}^2 
\int^{Q_{ECEC}+\epsilon_{b_1}+\epsilon_{b_2}}_{0}
\left( \left< K_N \right> ^2 +\left< L_N \right> ^2 +\left< K_N \right> \left< L_N \right> \right) \omega_1^2 \omega_2^2 d\omega_1.
\end{equation}
\end{widetext}

\begin{figure}[ht!]
\includegraphics[width=1.00\linewidth ]{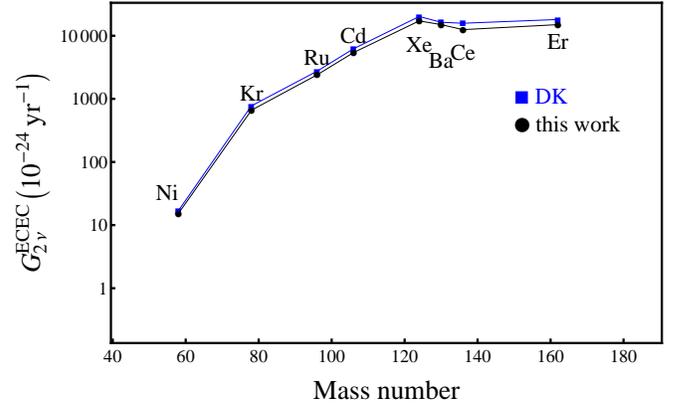} 
\caption{\label{2nuecec}(Color online) Phase space factors $G_{2\nu}^{ECEC}$ in units $(10^{-24}$ yr$^{-1})$. The label "DK" refers to the results obtained by Doi and Kotani \cite{doitwoneutrino} using approximate electron wave functions. The figure is in semilogarithmic scale.}
\end{figure}

In this case, in the definition of $\left< K_N \right>$ and $\left< L_N \right>$ in Eq. (\ref{closure}), $\epsilon_1=\epsilon_{e_1}=-(m_ec^2-e_{b_1})$ is the energy of the first captured electron and $\epsilon_2=\epsilon_{e_2}=-(m_ec^2-e_{b_2})$ is the energy of the second captured electron. The values obtained are listed in Table~\ref{table2} column 8 where they are  compared with previous calculations (columns 9 and 10), and in Fig.~\ref{2nuecec}. From $G_{2\nu}^{ECEC}$ we can calculate:\\
(i) The half-life
\begin{equation}
\label{t1/2_2necec}
\left[ \tau^{2\nu}_{1/2}\right]^{-1}=G^{EC EC}_{2\nu}g_A^4 \left| m_ec^2 M^{(2\nu)} \right| ^2.
\end{equation}

While in the case of $\beta^+\beta^+$ and $EC\beta^+$ decay all three calculations agree within a factor of $\sim 1.5$, in the case of $ECEC$ decay, the calculation reported in the book of Boehm and Vogel \cite{boehm}, disagrees with other two by a factor of approximately 4. The origin of this discrepancy is not clear. The values in Table~\ref{table2} have been converted to units yr$^{-1}$ using the same procedure in all three cases, $\beta^+\beta^+$, $EC\beta^+$, and $ECEC$. Since apart from the factor of 4, the behavior with mass number of $G_{2\nu}^{ECEC}$ in \cite{boehm} is the same as in the other two calculations, it may be simply due to a different definition of $G_{2\nu}^{ECEC}$. Note that the scale in Fig.~\ref{2nuecec}, $10^{-24}$ yr$^{-1}$, for $2\nu ECEC$ is very different from that for $2\nu\beta^+\beta^+$, $10^{-29}$ yr$^{-1}$, in Fig.~\ref{2nuGcompbb} due to a much larger $Q$-value


\subsection{Neutrinoless modes}
As discussed in Ref.~\cite{barea12b}, several scenarios of neutrinoless double beta decay have been considered, most notably, light neutrino exchange, heavy neutrino exchange, and Majoron emission. After the discovery of neutrino oscillations, attention has been focused on the first scenario and the mass mode, where the transition operator is proportional to $\left\langle m_{\nu}\right\rangle/m_e$. In this article we present phase-space factors for the mass mode. Phase-space factors associated with the other modes, called $\left\langle \lambda \right\rangle$ and $\left\langle \eta \right\rangle$ in Ref.~\cite{tom91}, will form the subject of a subsequent publication.

\subsubsection{$0\nu \beta^+\beta^+$ decay}
The equations for $0\nu\beta^+\beta^+$ decay are exactly the same as for $0\nu\beta^-\beta^-$ decay described in \cite{kotila12} where now $\epsilon_1$ is the energy of the first positron, $\epsilon_1=\epsilon_{p_1}$, and $\epsilon_2$ is the energy of the second positron, $\epsilon_2=\epsilon_{p_2}$. There are also here two quantities $G_{0\nu}^{(0)}$ and $G_{0\nu}^{(1)}$ in units of yr$^{-1}$ from which one can obtain:\\
(i) The half-life
\begin{equation}
\label{t1/2_0nbb}
\left[ \tau^{0\nu}_{1/2}\right]^{-1}=G^{\beta^+\beta^+}_{0\nu}g_A^4 \left|\frac{\left\langle m_{\nu}\right\rangle}{m_e}\right|^2\left| M^{(0\nu)} \right| ^2.
\end{equation}
(ii) The differential decay rate
\begin{equation}
\frac{dW_{0\nu}}{d\epsilon_{p_1}}={\cal N}_{0\nu}\ln 2\frac{dG^{\beta^+\beta^+}_{0\nu}}{d\epsilon_{p_1}},
\end{equation}
where ${\cal N}_{0\nu}=g_A^4 \left| M^{(0\nu)} \right|^2$.\\
Both the half-life and the differential decay rate, are given in terms of $G_{0\nu}^{(0)}\equiv G_{0\nu}^{\beta^+\beta^+}$.\\
(iii) The angular correlation between the two positrons
\begin{equation}
\alpha(\epsilon_{p_1})=\frac{dG^{(1)}_{2\nu}/d\epsilon_{p_1}}{dG^{(0)}_{2\nu}/d\epsilon_{p_1}}.
\end{equation}

\begin{figure}[ht!]
\includegraphics[width=1.00\linewidth ]{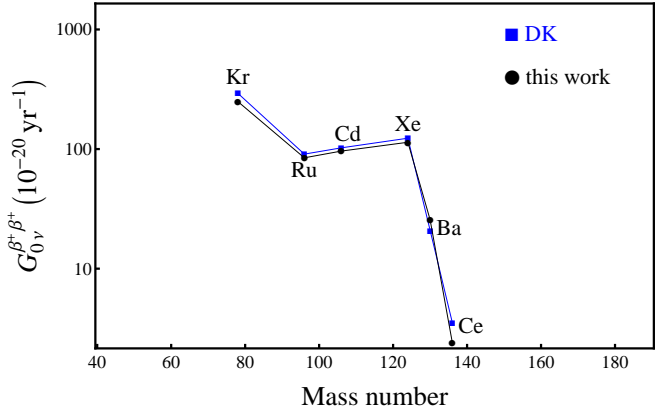} 
\caption{\label{0nubb}(Color online) Phase space factors $G_{0\nu}^{\beta^+\beta^+}$ in units $(10^{-20}$ yr$^{-1})$. The label "DK" refers to the results obtained by Doi and Kotani \cite{doitwoneutrino} using approximate electron wave functions. The figure is in semilogarithmic scale.}
\end{figure}

\begin{ruledtabular}
\begin{center}
\begin{table}[ht!]
\begin{tabular}{lcccccc}
   		&\multicolumn{3}{c}{$G_{0\nu}^{\beta^+\beta^+}(10^{-20}yr^{-1})$}		&\multicolumn{3}{c}{$G_{0\nu}^{EC\beta^+}(10^{-18}yr^{-1})$}		\\\cline{2-4} \cline{5-7}
\T Nucleus   	&This work 	&DK 	&BV 	&This work &DK &BV \\
\hline
&\multicolumn{6}{c}{\T \textbf{$\beta^+\beta^+$, $EC\beta^+$ and $ECEC$ allowed}}\\

$^{78}$Kr	&250		&293&	59.4	&6.37 		&7.11&	\\
$^{96}$Ru	&84.5		&90.7	&12.2	&9.62 		&10.8&	\\
$^{106}$Cd	&96.2		&102	&14.5	&13.0 		&14.7&	\\
$^{124}$Xe	&114		&123	&18.1	&19.7 		&22.9&	\\
$^{130}$Ba	&25.7		&21.0	&1.67	&17.6 		&19.8&	\\
$^{136}$Ce	&2.42		&3.55	&0.175	&15.3	 	&18.7&	\\
&\multicolumn{6}{c}{\T \textbf{$EC\beta^+$ and $ECEC$ allowed}}\\								

$^{50}$Cr		&		&	&&0.0887	&	&	\\
$^{58}$Ni		&		&	&&1.21	&1.30&	 	\\
$^{64}$Zn		&		&	&&0.0507	&	&	\\
$^{74}$Se		&		&	&&0.230	&	&\\
$^{84}$Sr		&		&	&&1.94	&	&		\\
$^{92}$Mo		&		&	&&1.92	&	&		\\
$^{102}$Pd		&		&	&&0.287	&	&\\
$^{112}$Sn		&		&	&&5.20	&	&\\
$^{120}$Te		&		&	&&3.92	&	&\\
$^{144}$Sm		&		&	&&8.11	&	&\\
$^{156}$Dy		&		&	&&15.2	&	&\\
$^{162}$Er		&		&	&&12.9	&15.8&\\
$^{168}$Yb		&		&	&&4.23	&	&\\
$^{174}$Hf		&		&	&&0.0272	&	&\\
$^{184}$Os		&		&	&&7.04	&	&\\
$^{190}$Pt		&		&	&&5.57	&	&\\
							

\end{tabular}
\caption{\label{table3}Phase space factors $G_{0\nu}$ obtained using screened exact finite size Coulomb wave functions. For comparison value of Doi and Kotani \cite{doineutrinoless} an Boehm and Vogel \cite{boehm} are also shown. These are extracted from \cite{doineutrinoless} and \cite{boehm} by removing $g_A^4$ and converting to yr$^{-1}$ units.}
\end{table}
\end{center}
\end{ruledtabular}
The values of $G_{0\nu}^{\beta^+\beta^+}$ are shown in Table~\ref{table3} column 2 where they are compared with previous calculations (column 3 and 4), and in Fig.~\ref{0nubb}. In this case, our calculation and that of \cite{doineutrinoless} disagree with the calculation reported in \cite{boehm} by a larger factor.

\begin{figure*}[cbt!]
\includegraphics[width=0.66\linewidth ]{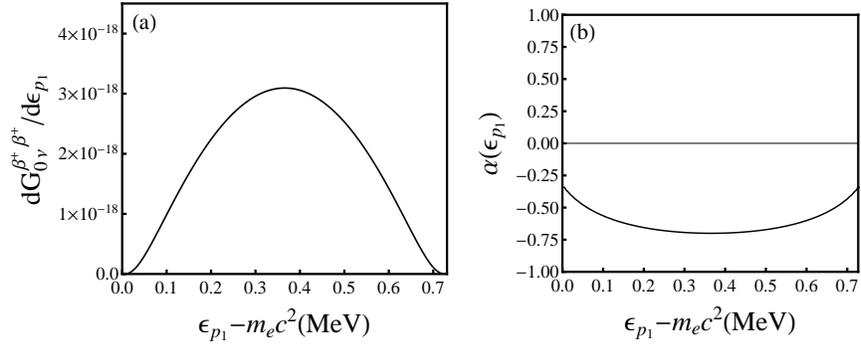} 
\caption{\label{cd0nu}Single positron spectra (panel a), and angular correlations between two outgoing positrons (panel b) for the $^{106}$Cd $\rightarrow ^{106}$Pd $0\nu\beta^+\beta^+$-decay. The scale in the left  should be multiplied by ${\cal N}_{0\nu}$ when comparing with experiment.}
\end{figure*}

We also have available upon request the single electron spectra and angular correlation for all $0\nu\beta^+\beta^+$ nuclei in Table~\ref{table3}. An example, $^{106}$Cd decay, is shown in Fig.~\ref{cd0nu}.

\subsubsection{$0\nu EC\beta^+$ decay}
\begin{figure}[ht!]
\includegraphics[width=1.00\linewidth ]{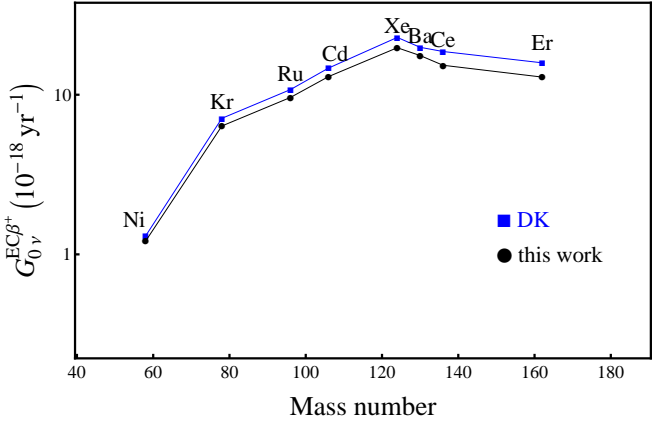} 
\caption{\label{0nuecb}(Color online) Phase space factors $G_{0\nu}^{EC\beta^+}$ in units $(10^{-18}$ yr$^{-1})$. The label "DK" refers to the results obtained by Doi and Kotani \cite{doineutrinoless} using approximate electron wave functions. The figure is in semilogarithmic scale.}
\end{figure}

In case of neutrinoless positron emitting electron capture, the energy of the emitted positron is fixed and the equation for PSF reads \cite{doineutrinoless}:
\begin{equation}
\begin{split}
G_{0\nu}^{EC\beta^+}&=\frac{1}{4R^2}\frac{2}{\ln2}\frac{(G\cos\theta)^4}{4\pi^3}(\hbar c^2)(m_ec^2)^5
\\ &\times
 \sum_{i=0,1}
 {\cal B}_{i,-1}^2
  \left[ (g_{-1}(\epsilon_p ,R))^2+(f_{1}(\epsilon_p ,R))^2\right] p_pc\epsilon_p.
\end{split}
\end{equation}
The PSF are in units yr$^{-1}$, and from those we can calculate:\\
(i) The half-life
\begin{equation}
\label{t1/2_0necb}
\left[ \tau^{0\nu}_{1/2}\right]^{-1}=G^{EC\beta^+}_{0\nu}g_A^4 \left|\frac{\left\langle m_{\nu}\right\rangle}{m_e}\right|^2\left| M^{(0\nu)} \right| ^2.
\end{equation}
The values obtained are listed in Table ~\ref{table3} column 5 where they are compared with previous calculations (column 8) and in Fig.~\ref{0nuecb}. In this case only calculations of \cite{doineutrinoless} are available.

\section{\label{secterr}Estimate of the error} 
Two main sources of error in the evaluation of the phase space factors are the $Q$-values and the nuclear radius, $R$. We have taken for the atomic mass $M(A,Z)$ the available experimental values with errors shown in Table~\ref{Qval}. The more accurate values used in this table account for some of the differences between our calculated values and those of \cite{doitwoneutrino, doineutrinoless, boehm} obtained with older values of the atomic masses. We estimate the error here as a multiple of $\delta Q/Q$, where $\delta Q$ is the error in $Q$. The nuclear radius enters the calculation in various ways, most notably the evaluation of the positron wave functions at the nucleus $g_{-1}(R)$ and $f_{1}(R)$ and for $EC\beta^+$ and $ECEC$, the electron probability, ${\cal B}_{n',\kappa}^2$. We have taken $R=r_0A^{1/3}$ with $r_0=1.2$fm. An estimate of the error here is obtained as in single-$\beta^+$ decay and single EC \cite{buhring} by adjusting $r_0$ for each nucleus, $A,Z$, using the experimental value $\langle r^2 \rangle_{exp}$ from electron scattering. Finally, another uncertainty is introduced by the average excitation energy, $\langle E_N \rangle$. In \cite{kotila12} we estimated this uncertainty which affects only the $2\nu$ processes by comparing the results of the calculation with $\langle E_N \rangle =1.12A^{1/2}$MeV with that of the single state dominance model, for example in $^{106}$Cd decay, the ground state of the intermediate $^{106}$Ag nucleus is $1^+$,  giving $E_N=E_{1_1^+}=0.0$MeV. The resulting error is of the order of few percent. Screening corrections play a minor role and do not introduce a large error. The situation is summarized in Table~\ref{unc}.
\begin{ruledtabular}
\begin{center}
\begin{table}[ht!]
\begin{tabular}{lcc}
\T$2\nu\beta^+\beta^+$ 	&$Q$-value	&$10 \times \delta Q/Q $\\
			&Radius 	&$1.0\%$ \\
			&Screening 	&$0.10\%$\\
			&$\langle E_N \rangle$ &model dependent ($\leq 1\%$)\\
$0\nu\beta^+\beta^+$ 	&$Q$-value	&$4 \times \delta Q/Q $\\
			&Radius 	&$9\%$ \\
			&Screening 	&$0.10\%$\\
			&$\langle E_N \rangle$ &-\\
\cr \hline 
\T$2\nu EC\beta^+$ 	&$Q$-value	&$8 \times \delta Q/Q $\\
			&Radius 	&$1.0\%$ \\
			&Screening 	&$0.10\%$\\
			&$\langle E_N \rangle$ &model dependent ($\leq 1\%$)\\
$0\nu EC\beta^+$ 	&$Q$-value	&$2 \times \delta Q/Q $\\
			&Radius 	&$9\%$ \\
			&Screening 	&$0.10\%$\\
			&$\langle E_N \rangle$ &-\\
\cr \hline 
\T$2\nu ECEC$ 	&$Q$-value	&$6 \times \delta Q/Q $\\
			&Radius 	&$1.0\%$ \\
			&Screening 	&$0.10\%$\\
			&$\langle E_N \rangle$ &model dependent ($\leq 1\%$)\\
\end{tabular}
\caption{\label{unc}The estimate of uncertainties introduced to phase space factors due to different input parameters.}
\end{table}
\end{center}
\end{ruledtabular}

\section{Use of phase space factors}
\begin{ruledtabular}
\begin{center}
\begin{table*}[cbt!]
\begin{tabular}{lccc}
Nucleus   &$G_{2\nu}^{ECEC}(10^{-21}$yr$^{-1})$  	&$\tau_{1/2}^{2\nu ECEC}(10^{21}$ yr) exp\footnotemark[1] &$|M^{\rm{eff}}_{2\nu ECEC}|$ \cr \hline
\T
$^{130}$Ba	&15.0  	&$2.2 \pm 0.5$		&$0.174\pm0.017$	\\
\end{tabular}
\footnotetext[1]{Ref.~\cite{53}}
\caption{\label{tablemeff}Experimental $2\nu ECEC$ half-lives and the corresponding effective nuclear matrix elements $|M^{\rm{eff}}_{2\nu ECEC}|$. 
}
\end{table*}
\end{center}
\end{ruledtabular}

The main use of PSFs discussed in this paper is to calculate half-lives for the $\beta^+\beta^+$, $EC\beta^+$ and $ECEC$ decay, by combining them with a calculation of the NME, the only constraints being that the NME are defined in a way consistent with Eq. (\ref{t1/2_2nbb}), $2\nu\beta^+\beta^+$, Eq. (\ref{t1/2_2necb}), $2\nu EC\beta^+$, Eq. (\ref{t1/2_2necec}), $2\nu ECEC$, Eq. (\ref{t1/2_0nbb}), $0\nu\beta^+\beta^+$, and Eq. (\ref{t1/2_0necb}), $0\nu EC\beta^+$. The calculation of half-lives with matrix elements obtained from IBM-2 will be presented in forthcoming publication \cite{barprep12}. Here we use the calculation of PSF to extract the dimensionless quantity $g_A^4|(m_ec^2)M_{2\nu}|^2=|M_{2\nu}^{\rm{eff}}|^2$as done in the case of $\beta^-\beta^-$ decay reported in \cite{kotila12}.

For two neutrino double positron decay and competing modes the only positive experimental half-life result is from a geochemical
experiment in $^{130}$Ba \cite{53}: $T_{1/2}^{2\nu}~=~(2.2~\pm~0.5)~\times~10^{21}$yr. In geochemical experiments, it is not possible to disentangle the
different modes, but in Ref.~\cite{barabash10} this value is believed to be for the $2\nu ECEC$ process, because other modes are strongly suppressed. Our calculations in Table~\ref{table2} support this statement and we thus take $T_{1/2}^{2\nu ECEC} = (2.2 \pm 0.5) \times 10^{21}$yr in $^{130}$Ba, from which we extract the value of $|M_{2\nu ECEC}^{\rm{eff}}|$ in Table~\ref{tablemeff}.

\begin{figure}[ht!]
\includegraphics[width=1.00\linewidth ]{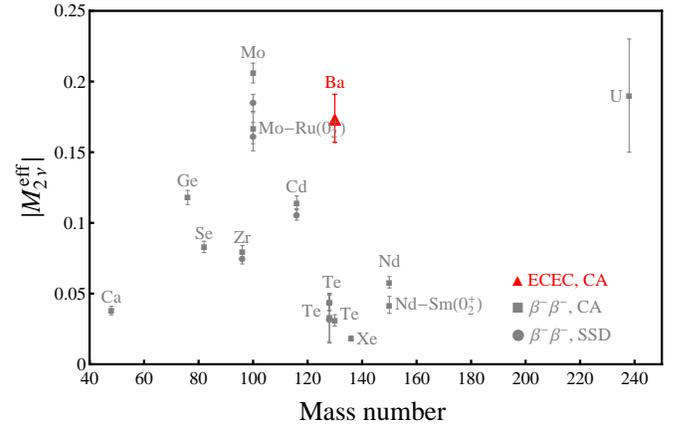} 
\caption{\label{Mefffig}(Color online) Effective nuclear matrix elements $|M^{\rm{eff}}_{2\nu}|$ extracted from the experimental $2\nu\beta\beta$ half-lives as a function of mass number. The red triangle is the value extracted from $\tau_{1/2}^{2\nu ECEC}$. For comparison the grey dots and squares represent the values extracted from $\beta^-\beta^-$ experiments.}
\end{figure}

We note that the value we extract is comparable to the values extracted from $2\nu\beta^-\beta^-$ decay in \cite{kotila12}. To emphasize this point we show in Fig.~\ref{Mefffig} the newly extracted value in comparison with all others.

\section{Conclusions}
In this article, we have reported a complete and improved calculation of phase space factors for $2\nu\beta^+\beta^+$, $2\nu EC\beta^+$ and $2\nu ECEC$, as well as $0\nu\beta^+\beta^+$ and $0\nu EC\beta^+$ double-$\beta$ decay modes, including half-lives, single positron spectra, summed positron spectra, and positron angular correlations, to be used in connection with the calculation of nuclear matrix elements. Apart from their completeness and consistency of notation, we have improved the calculation by using exact Dirac wave function with finite nuclear size and electron screening. The program for calculation of phase space factors has been set up in such a way that additional improvements may be included if needed (P-wave contribution, finite extent of nuclear surface, etc.) and that it can be used in connection with the closure approximation, the single state dominance hypothesis and the calculation with sum over individual states. 

\begin{acknowledgments}
This work was performed in part under the US DOE Grant DE-FG-02-91ER-40608. We wish to thank 
K. Zuber for stimulating discussions.
\end{acknowledgments}


\begin{thebibliography}{99}
\bibitem{barea09} J. Barea and F. Iachello, Phys. Rev. C \textbf{79}, 044301 (2009).
\bibitem{kotila12}J. Kotila and F. Iachello, Phys. Rev. C \textbf{85}, 034316 (2012).
\bibitem{barea12}J.\ Barea, J.\ Kotila and F.\ Iachello, Phys. Rev. Lett. \textbf{109}, 042501 (2012).
\bibitem{barea12b}J.\ Barea, J.\ Kotila and F.\ Iachello, Phys. Rev. C \textbf{87}, 014315 (2013).
\bibitem{zuber} K. Zuber, private communication.
\bibitem{barprep12} J.\ Barea, J.\ Kotila and F.\ Iachello (private communication).
\bibitem{primakoff1} H. Primakoff and S.P. Rosen, Rep. Prog. Phys. \textbf{22}, 121 (1959).
\bibitem{primakoff2} H. Primakoff and S.P. Rosen, Proc. Phys. Soc. \textbf{78}, 464 (1961).
\bibitem{haxton} W. C. Haxton and G.J. Stephenson Jr., Prog. Part. Nucl. Phys. \textbf{12}, 409 (1984).
\bibitem{jotain} J. D. Vergados, Phys. Lett. B \textbf{109}, 96 (1982).
\bibitem{jotain2} J. D. Vergados, Nucl. Phys. B \textbf{218}, 109 (1983).
\bibitem{doitwoneutrino} M. Doi and T. Kotani, Prog. Theor. Phys. \textbf{87}, 1207 (1992).
\bibitem{doineutrinoless} M. Doi and T. Kotani, Prog. Theor. Phys. \textbf{89}, 139 (1993).
\bibitem{doi} M. Doi, T. Kotani, and E. Takasugi, Prog. Theor. Phys. suppl. \textbf{83}, 1 (1985).

\bibitem{boehm} F. Boehm and P. Vogel, \textit{Physics of Massive Neutrinos} (Cambridge University Press, New York, 1992).
\bibitem{single} B.H. Wilkinson and B.E.F. Macefield, Nucl. Phys. \textbf{A232}, 58 (1974).
\bibitem{bambynek}W. Bambynek \textit{et al.}, Rev. Mod. Phys. \textbf{49}, 77–221 (1977).

\bibitem{rose} M.E. Rose, \textit{Relativistic Electron Theory} (Wiley, New York, 1961).
\bibitem{sal95} F. Salvat, J. Fernadez-Varea, and W. Williamson Jr., Comp. Phys. Comm. \textbf{90}, 151 (1995).
\bibitem{wil70} D. H. Wilkinson, Nucl. Phys. A \textbf{150}, 478 (1970).
\bibitem{buhring} W. B\"uhring, Nucl. Phys. \textbf{61}, 110 (1965).
\bibitem{esp02} S. Esposito, Am. J. Phys. \textbf{70}, 852 (2002).
\bibitem{tom91} T. Tomoda, Rep. Prog. Phys. \textbf{54}, 53 (1991).
\bibitem{nudat}NuDat 2.6, http://www.nndc.bnl.gov/nudat2/
\bibitem{prc83}S. Eliseev \textit{et al.}, Phys. Rev. C \textbf{83}, 038501 (2011). 
\bibitem{prc84}M. Goncharov \textit{et al.}, Phys. Rev. C \textbf{84}, 028501 (2011).
\bibitem{plb697}V.S. Kolhinen \textit{et al.}, Phys. Lett. B \textbf{697} 116 (2011).
\bibitem{plb684}V.S. Kolhinen \textit{et al.}, Phys. Lett. B \textbf{684} 17 (2010).
\bibitem{prl103}S. Rahaman \textit{et al.}, Phys. Rev. Lett. \textbf{103}, 042501 (2009). 
\bibitem{prc80}N. D. Scielzo \textit{et al.}, Phys. Rev. C \textbf{80}, 025501 (2009).
\bibitem{prc86}C. Smorra \textit{et al.}, Phys. Rev. C \textit{86}, 044604 (2012).
\bibitem{arxiv1201}C. Smorra \textit{et al.}, arXiv:1201.4942 (2012).
\bibitem{prl106}S. Eliseev \textit{et al.}, Phys. Rev. Lett. \textbf{106}, 052504 (2011).
\bibitem{prl107}S. Eliseev \textit{et al.}, Phys. Rev. Lett. \textbf{107}, 152501 (2011).
\bibitem{arxiv111}C. Droese \textit{et al.}, arXiv:1111.6377 (2011).
\bibitem{53} A. P. Meshik, C. M. Hohenberg, O. V. Pravdivtseva, and Y. S.
Kapusta, Phys. Rev. C \textbf{64}, 035205 (2001).
\bibitem{barabash10} A. S. Barabash, Phys. Rev. C \textbf{81}, 035501 (2010).


\end{thebibliography}
\end{document}